\documentclass[11pt]{article}
\usepackage{amsthm, amsfonts}
\usepackage{fullpage}
\usepackage{amscd,amsmath,amssymb}

\newtheorem{lemma}{Lemma}
\newtheorem{proposition}{Proposition}
\newtheorem{corollary}{Corollary}

\newcommand{\phip}{\phi^\dagger}
\newcommand{\Bip}{B^\dagger}
\newcommand{\la}{\lambda}
\newcommand{\La}{\Lambda}
\newcommand{\eps}{\varepsilon}
\newcommand{\be}{\begin{equation}}
\newcommand{\ee}{\end{equation}}

\newcommand{\B}{{\cal B}}

\title{Quantum inverse scattering method for the $q$-boson model
and symmetric functions}
\author{N.~V.~Tsilevich\footnote{St.~Petersburg Department of
Steklov Institute of Mathematics.
E-mail: {\tt natalia@pdmi.ras.ru}. Supported by CRDF grant RUM1-2622-ST-04, 
RFBR grant 05-01-00899, and the President of Russian Federation grant for
Support of Leading Scientific Schools NSh.-2251.2003.1.}}
\date{}

\begin{document}
\maketitle

\section{Introduction}

The {\it $q
$-boson model} (see, e.g., \cite{BBT, BIK2})
describes a strongly correlated 
exactly solvable one-dimensional boson system on a finite
chain which is of importance in several branches
of modern physics, such as solid state physics and quantum nonlinear optics.
The corresponding $q$-{\it boson} (or $q$-{\it oscillator}) algebra \cite{KD}
is closely related to the quantum algebra $sl_q(2)$ \cite{Kulish}.
The particular case $q=0$ of the $q$-boson model,
which is especially easy to investigate, is called the 
{\it phase model} \cite{BIK1, BIK2, Bogol}.

The purpose of this paper is to show that the {\it quantum inverse scattering
method} (QISM) \cite{KBI} for the $q$-boson model has
a nice interpretation in terms of the

{\it algebra of symmetric functions} \cite{Mac}.
The starting point for our approach was the paper \cite{Bogol} 
by N.~M.~Bogoliubov,
who showed that the phase model is closely related to enumeration of plane
partitions.

Starting from the more simple case of the phase model,
we construct its realization in the algebra of symmetric 
functions $\La$ by associating with a basis Fock vector $\psi_{n_0,\ldots,n_M}$
with occupation numbers $n_0,\ldots,n_M$ the Schur function $s_\la(x)$
corresponding to the Young diagram $\la$ that has $n_j$ rows of length $j$
(for details, see Sec.~\ref{sec:symmphase}). It turns out that under
this realization, the creation operator
$B(u)$ of QISM
coincides (up to a scalar factor) with the operator of multiplication
by the (truncated) generating function $H_M(u^2)=\sum_{k=0}^Mu^{2k}h_k$
of the complete homogeneous symmetric functions $h_k$,
and the annihilation operator $C(u)$ is essentially
the adjoint operator $H_M^\perp(u^{-2})$
with respect to the standard scalar product in $\La$. 
This allows us, in particular, to apply the machinery of 
symmetric functions to
immediately obtain an expansion of the 
wave function in terms of the basis Fock vectors;
the coefficients of this expansion are given by Schur functions.
Besides, we can easily
find the $M\to\infty$ limit of the regularized creation and annihilation 
operators.

On the other hand, we can use
this interrelation between the phase model
and symmetric functions in the opposite
direction: for example, using
the commutation relations for $B
(u)$ and $C(u)$ given
by QISM ($R$-matrix), we can
obtain commutation
relations for $H_M(u)$ and $H_M^\perp(u)$ in  the subspace $\La_M$
of $\La$ spanned by Schur functions whose diagrams have at most $M$ columns
(they are more involved than the commutation relation for the 
operator of multiplication
by the full generating function 
$H(u)=\sum_{k=0}^\infty u^{k}h_k$ and its adjoint
$H^\perp(u)$ in the whole algebra $\La$).

We also establish a relation of the operators arising in 
QISM for the phase model to the
vertex operator formalism used by A.~Okounkov and N.~Reshetikhin \cite{OkResh}
for computing the correlation functions of three-dimensional Young diagrams
(plane partitions).
It turns out that the vertex operators from \cite{OkResh}
are the same operators $H(u)$ and $H^\perp(u)$, i.e., the $M\to\infty$ limits
of the regularized creation and annihilation operators
of the phase model. However, if we want to study three-dimensional Young
diagrams contained in a {\it box},
the approach of \cite{OkResh} fails,
while the method of \cite{Bogol}, based on QISM for the phase model,
allows one to compute the partition and  correlation functions
for three-dimensional diagrams in a box, since, as noted above, 
it allows one to obtain commutation relations for the ``truncated''
operators.

A similar scheme can be implemented for the
general $q$-boson model. In this case, one should use a
generalization of the Schur functions, namely, the
{\it Hall--Littlewood functions} $P_\la(x;q^2)$ 
(which are specialized to $s_\la(x)$ for $q=0$). In particular, 
the wave functions
of the $q$-boson model are expressed in terms of the Hall--Littlewood functions, and the
creation operator is essentially
the operator of multiplication by the generating function
$Q_M(u)=\sum_{k=0}^Mu^{2k}q_k$ (see Sec.~\ref{sec:q}).

\section{Phase model and Schur functions}
\label{sec:phase}

\subsection{Phase model}
\label{sec:defphase}

Consider the algebra
generated by three operators $\phi,\phip, N$ with
commutation relations
\be
[N,\phi]=-\phi,\qquad [N,\phip]=\phip,\qquad [\phi,\phip]=\pi, 
\label{crphase}
\ee
where $\pi$ is the vacuum projection.
This algebra can be realized in the 
one-dimensional (i.e., having one-dimensional $n$-particle subspaces) Fock 
space, the operators  $\phi,\phip$, and $N$ acting as the
phase operators and the number of particles operator, respectively:
$$
\phip|n\rangle=|n+1\rangle, \qquad
\phi|n\rangle=|n-1\rangle,\; \phi|0\rangle=0,\qquad
N|n\rangle=n|n\rangle,
$$
where $|n\rangle$ is the (normalized) $n$-particle Fock vector
(in particular, $|0\rangle$ is the vacuum vector).
Thus $\phi$ is an isometry
(one-sided shift), and we have
$$
\phi\phip=1, \qquad
\phip\phi=1-\pi,
$$
where $\pi|0\rangle=|0\rangle$, $\pi|n\rangle=0$ for $n\ge1$.

Now fix a positive integer $M$ (the number of sites)
and consider the tensor product 
${\cal F}={\cal F}_0\otimes{\cal F}_1\otimes\ldots\otimes{\cal F}_{M}$
of $M+1$ copies ${\cal F}_i$, $i=0,\ldots,M$, of the one-dimensional Fock space.
Denote by $\phi_i,\phip_i, N_i$ the operators that act as
$\phi,\phip, N$, respectively, in the $i$th space, and identically
in the other spaces; for example,
$$
\phi_i=I_1\otimes\ldots\otimes I_{i-1}\otimes\phi\otimes I_{i+1}\otimes\ldots
\otimes I_{M},
$$
where $I_j$ is the identity operator in ${\cal F}_j$.
Thus the operators commute at different sites and
satisfy the commutation relations (\ref{crphase}) at the same site.

The Hamiltonian of the phase model has the form
$$
H=-\frac12\sum_{n=0}^M(\phip_n\phi_{n+1}+\phi_n\phip_{n+1}-2N_n),
$$
with the periodic boundary conditions: $M+1\equiv 1$.

Following the quantum inverse scattering method \cite{KBI},
consider the {\it $L$-matrix} 
$$
L_n(u)=\left(\begin{array}{cc}
u^{-1}I & \phip_n\\
\phi_n & uI
\end{array}\right), \qquad n=0,1,{\ldots} ,M,
$$
where $u$ is a scalar parameter and $I$ is the identity 
operator in $\cal F$. This $L$-matrix, for every
$n=0,1,{\ldots} ,M$, satisfies the bilinear equation
\be
R(u,v)(L_n(u)\otimes L_n(v))=(L_n(v)\otimes L_n(u))R(u,v),
\label{Lbil}
\ee
with the $4\times 4$ $R$-matrix $R(u,v)$ given by
\be
R(u,v)=\left(\begin{array}{cccc}
f(v,u) & 0 & 0 & 0\\
0 & g(v,u) & 1 & 0\\
0 & 0& g(v,u) & 0\\
0 & 0 & 0 & f(v,u)
\end{array}\right),
\label{R-phase}
\ee
where 
\be
f(v,u)=\frac{u^2}{u^2-v^2},\qquad g(v,u)=\frac{uv}{u^2-v^2}.
\label{fg-phase}
\ee

The {\it monodromy matrix} is defined as
$$
T(u)=L_M(u)L_{M-1}(u)\ldots L_0(u)\qquad (\mbox{matrix product}).
$$
It satisfies the bilinear equation with the same $R$-matrix (\ref{R-phase}),
(\ref{fg-phase}):
\be
R(u,v)(T(u)\otimes T(v))=(T(v)\otimes T(u))R(u,v).
\label{Tbil}
\ee

Let 
$$
T(u)=\left(\begin{array}{cc}
A(u) & B(u)\\
C(u) & D(u)
\end{array}\right).
$$
The matrix entries $A(u), B(u), C(u), D(u)$ of the monodromy
matrix $T(u)$ act in the space $\cal F$. Denoting by
$\hat N
=N_0+\ldots+N_m$ the operator of the total number of particles,
we have 
$$
\hat N B(u)=B(u)(\hat N+1),\qquad
\hat N C(u)=C(u)(\hat N-1),
$$
so that $B(u)$ is a creation operator and $C(u)$
is an annihilation operator. The operators $A(u)$ and $C(u)$
do not change the number of particles.

Denote by $|0\rangle_j$ the vacuum vector in ${\cal F}_j$
and by
$|0\rangle=\bigotimes_{j=0}^M |0\rangle_j$
the total vacuum vector in
${\cal F}$.
Consider $N$-particle state vectors of the form
$$
|\Psi_N(u_1,\ldots,u_N)\rangle=\prod_{j=1}^NB(u_j)|0\rangle
$$
(according to the algebraic Bethe Ansatz (see \cite{KBI}),
the eigenfunctions of the Hamiltonian are precisely of this form).
We are interested in calculating the expansion of these vectors in
terms of basis $N$-particle vectors
\be
\psi_{n_0,\ldots,n_M}=\bigotimes_{j=0}^M|n_j\rangle_j,\qquad n_0+\ldots+n_M=N,
\label{Npart}
\ee
where $|n_j\rangle_j=(\phip)^n|0\rangle_j$ is the $n_j$-particle vector in the
$j$th Fock space ${\cal F}_j$; the numbers $n_k$ are called the
{\it occupation numbers} of the vector (\ref{Npart}).

\subsection{Realization of the phase model in the algebra of symmetric functions}
\label{sec:symmphase}

Necessary background on symmetric functions can be
found in \cite{Mac}.

Given a basis vector (\ref{Npart}) 
with occupation numbers $n_0,\ldots,n_M$, we associate with it the
Young diagram $\la=1^{n_1}2^{n_2}\ldots$ that has $n_j$ rows of 
length $j$, $j=1,\ldots, M$, and the
corresponding Schur function\footnote{Recall that
the Schur functions form
a basis in the algebra of symmetric functions $\La$.} $s_\la$:
\be
\bigotimes_{j=0}^M|n_j\rangle_j\leftrightarrow s_\la,\qquad \la=1^{n_1}2^{n_2}\ldots.
\label{corr}
\ee

\smallskip\noindent{\bf Remark.} Bearing in mind this correspondence
on the one hand and the standard realization of the Fock space
in the algebra of symmetric functions (see, e.g., \cite[Ch.~14]{Kac})
on the other hand, it
is natural to call the $j$th Fock space ${\cal F}_j$ the
{\it $j$-energy space}.
\smallskip

Note that the correspondence (\ref{corr}) does not
take into account the number $n_0$ of zero-energy particles. 
Thus (\ref{corr}) defines a realization
of the positive-energy space 
$\widehat{\cal F}={\cal F}_1\otimes\ldots\otimes{\cal F}_{M}$ 
in the algebra of symmetric functions $\La$, or, more exactly, 
in its subspace $\La_M$ generated by Schur functions $s_\la$ whose
diagrams have
at most $M$ columns (in other terms,
rows of length at most $M$). In view of the Jacobi--Trudi
identity \cite[I.3.4]{Mac}, one can determine this
subspace by choosing\footnote{This can be done,
because, as is well-known, the complete symmetric functions $h_k$
are algebraically independent in $\La$.}
the arguments of the symmetric functions
so that 
\be
h_{M+1}=h_{M+2}=\ldots=0
\label{spec}
\ee
(for details, see the proof of Proposition~\ref{prop:schurs}).

On the 
other hand, if we know the total number of particles $N$,
then we can recover the number of zero-energy particles in the basis vector
corresponding to $s_\la$
as $n_0=N-l(\la)$, where $l(\la)$ is the number of rows of the diagram $\la$. 
Consider the decomposition 
${\cal F}={\cal F}^0\oplus{\cal F}^1\oplus\ldots\oplus{\cal F}^N\oplus\ldots$
of the whole space  ${\cal F}$
into $N$-particle subspaces ${\cal F}^N$,
and let $\La_M^N$ be the space of symmetric functions corresponding
to ${\cal F}^N$. Note that the space $\La_M^N$
is spanned by Schur functions $s_\la$ whose diagrams $\la$ lie in the
$M\times N$ box, i.e., have
at most $N$ rows and at most $M$ columns. 
Thus the whole space ${\cal F}$ can be realized as the direct sum
${\cal F}\equiv\La^0_M\oplus\La^1_M\oplus\ldots\oplus\La^N_M\oplus\ldots$.

Since $B(u)$ is a creation 
operator, i.e., increases the number of particles by one, it
sends $\La^N_M$ to $\La^{N+1}_M$. Thus it suffices to study its 
action on the space $\widehat{\cal F}\equiv\La_M$, i.e.,
the operator ${\cal B}(u):=PB(u)P$,
where $P$ is the projection from $\cal F$ to $\widehat{\cal F}$
(``forgetting the zero-energy space'').

\begin{proposition} 
Let ${\cal B}(u)=u^{-M}\tilde{\cal B}(u)$. 
The operator $\tilde{\cal B}(u)$ acts in $\La_M$
as the operator of multiplication by $H_M(u^2)$,
where $H_M(t)=\sum_{k=0}^M t^kh_k$
is the (truncated) generating function of the complete
homogeneous symmetric functions $h_k$.
\label{prop:B}
\end{proposition}

\begin{proof}
It is easy to see that the operator $\tilde\B$ 
has the form $\tilde\B=\sum_{k=0}^M u^{2k}\B_k$. Thus it suffices to prove
that $\B_k$ is the operator of multiplication by the $k$th complete
symmetric function $h_k$. Denote  $\phi_j^{-1}=\phi_j$, $\phi_j^{0}=1$,
$\phi_j^{1}=\phip_j$.
Since
$$B(u)=\sum_{j_M,\ldots,j_1=1}^2 (L_M(u))_{1j_M} (L_{M-1}(u))_{j_Mj_{M-1}}\ldots
(L_0(u))_{j_12},
$$
we have
$$
\B_k=\sum_{\eps_M,\ldots,\eps_0}\phi_M^{\eps_M}\ldots\phi_1^{\eps_1}=
\sum_{\eps_M,\ldots,\eps_0}\B_{\eps_M,\ldots,\eps_0},
$$
where the sum is over all collections $\eps_j\in\{-1,0,1\}$, $j=0,{\ldots},M$, 
satisfying the following conditions:
(a) let $\eps_l$ be the highest nonzero element, i.e.,
$\eps_M=\ldots\eps_{l+1}=0$ and $\eps_l\ne0$; then $\eps_l=1$;
(b) $\eps_0\ne -1$; (c) adjacent elements do not have the same sign,
i.e., $\eps_{j+1}\eps_{j}\ne 1$ for every $j$; (d) $\sum_{j=1}^Mj\eps_j=k$. 
Obviously, $\B_{\eps_M,\ldots,\eps_0}$ sends a basis vector~(\ref{Npart})
to a basis vector.

In terms of Schur functions, we have 
$\phip_js_\mu=s_\la$, where the diagram $\la$ is obtained from $\mu$
by inserting a row of length $j$, and $\phi_j s_\mu=s_\la$, where $\la$ 
is obtained from $\mu$ by removing a row of length $j$ 
(with $\phi_j s_\mu=0$ if $\mu$ does not contain a row of length $j$).
Denote by $\nu'_i$ the length of the $i$th
column of a diagram $\nu$ and by $n_i(\nu)$ 
the number of rows of length $i$ in $\nu$. Then
$\nu'_i-\nu'_{i+1}=n_i(\nu)$. Now let $\B_{\eps_M,\ldots,\eps_0}s_\mu=s_\la$ and
set
$\theta'_i=\la'_i-\mu'_i$. Then $n_i(\la)=n_i(\mu)+\eps_i$, so that
$\theta'_i=\theta'_{i+1}+n_i(\la)-n_i(\mu)=
\theta'_{i+1}+\eps_i$, whence
$$
\theta'_M=\eps_M,\quad \theta'_{M-1}=\eps_M+\eps_{M-1},\;\ldots,\;
\theta'_1=\eps_M+\ldots+\eps_1.
$$
Now it follows from (a) that $\theta'_M=\ldots=\theta'_{l+1}=0$,
$\theta'_l=1$. Further, 
in view of (c), we have
$\theta'_i\in\{0,1\}$,
which means that $\la\supset\mu$ and 
the skew diagram $\la\setminus\mu$ contains
at most one cell in each column, i.e., is
a horizontal strip. Moreover, (d) implies $\sum\theta'_i=k$,
so that $\la\setminus\mu$ contains $k$ cells.
Denoting by ${\cal H}_k$
the set of horizontal $k$-strips, we obtain
$$
\B_ks_\mu=\sum_{\la:\,\la\setminus\mu\in{\cal H}_k}s_\la,
$$
whence, in view of the Pieri formula
\cite[I.5.16]{Mac}, $\B_ks_\mu=h_ks_\mu$.
\end{proof}

\smallskip\noindent{\bf Remark.} The 
truncated generating function $H_M(t)=\sum_{k=0}^Mt^kh_k$ can also be regarded 
as the full generating function $H(t)=\sum_{k=0}^\infty t^kh_k$ under the
specialization~(\ref{spec}): 
$H_M(t)=H(t)\left|_{h_{M+1}=h_{M+2}=\ldots=0}\right.$.

\begin{corollary}
The $M\to\infty$ limit of the regularized creation
operator $\tilde {\cal B}(u)=u^M{\cal B}(u)$ on the positive-energy subspace 
$\widehat{\cal F}$ is just 
the multiplication by $H(u^2)$ in the whole algebra
of symmetric functions $\La$.
\end{corollary}

Using the interpretation of the operator ${\cal B}(u)$ obtained in Proposition~\ref{prop:B},
we can easily find the required expansion of the $N$-particle
vector (\ref{Npart}) in terms of basis vectors.

\begin{proposition} 
The expansion of the $N$-particle
vector {\rm(\ref{Npart})} in terms of basis vectors is given by the formula
$$
|\Psi_N(u_1,\ldots,u_N)\rangle=
\sum_{\la}s_\la(u_1^2,\ldots, u_N^2)\bigotimes_{j=0}^M|n_j\rangle_j,
$$
where the sum is over Young diagrams $\la$ with at most $N$ rows
and at most $M$ columns.
\label{prop:schurs}
\end{proposition}
\begin{proof}
By the formula for the generating function
of complete symmetric functions \cite[I.2.5]{Mac},
$$
H(u^2)=\prod_{i}\frac1{1-u^2x_i}.
$$
Note that $\Psi_N(u_1,\ldots,u_N)\in{\cal F}^N$ and
identify ${\cal F}^N$ with $\La_M^N$ as described above. Observing 
that the vacuum vector corresponds to the unit function
$s_\emptyset\equiv1$ and using  Proposition~\ref{prop:B}, we obtain
\begin{eqnarray*}
|\Psi_N(u_1,\ldots,u_N)\rangle&=&\prod_{j=1}^NB(u_j)|0\rangle=
(u_1\ldots u_M)^{-M}\prod_{j=1}^N\tilde B(u_j)|0\rangle\\
&=&(u_1\ldots u_M)^{-M}\prod_j\prod_i\frac1{1-u_j^2x_i}.
\end{eqnarray*}
The well-known Cauchy identity \cite[I.4.3]{Mac} yields
$$
|\Psi_N(u_1,\ldots,u_N)\rangle=(u_1\ldots u_M)^{-M}
\sum_\la s_\la(u_1^2,\ldots,u_N^2)s_\la(x),
$$
which gives the desired formula in view of (\ref{corr}).
The restrictions on $\la$ are obtained as follows. 
First, a Schur function vanishes if the number of nonzero arguments
is less than the number of its rows. Thus $s_\la(u_1^2,\ldots,u_N^2)=0$
if $l(\la)>N$. On the other hand, by the Jacobi--Trudi identity
\cite[I.3.4]{Mac} we have $s_\la=\det(h_{\la_i-i+j})_{i,j=1}^n$,
where $n\ge l(\la)$. Thus we see that under the specialization (\ref{spec}),
the first row of this determinant, and hence $s_\la$,
vanishes if $\la_1> M$.
\end{proof}

\begin{lemma}
\label{l:ABCD}
The matrix entries of the monodromy matrix $T(u)$
are related by the following formulas:
$$
B(u)=uA(u)\phip_0,\qquad
C(u)=u^{-1}\phi_0A^\dagger(u^{-1}),\qquad
D(u)=\phi_0A^\dagger(u^{-1})\phip_0.
$$
\end{lemma}
\begin{proof}
Easy induction on $M$.
\end{proof}

In particular, setting ${\cal A}(u)=PA(u)P$, ${\cal C}(u)=PC(u)P$, and 
${\cal D}(u)=PD(u)P$, we have 
$$
{\cal A}(u)=u^{-1}{\cal B}(u),\qquad {\cal C}(u)={\cal B}^\dagger(u^{-1}),
\qquad {\cal D}(u)=u{\cal B}^\dagger(u^{-1}).
$$
It follows, for example, that in the realization
of the phase model in the algebra of symmetric functions, 
the annihilation operator has the following representation:
$$
{\cal C}(u)=u^M\tilde{\cal C}(u),\qquad
\tilde{\cal C}(u)=\tilde\B^\dagger(u^{-1})=H^\perp_M(u^{-2})=
\sum_{n=0}^M u^{-2n} h^{\perp,M}_n,
$$
where $h^{\perp, M}_n$ is the adjoint to the operator of multiplication 
by $h_n$ in the space $\La_M$
with the standard scalar product (with respect to which the Schur functions form
an orthonormalized basis). 
Note that $h^{\perp, M}_n$ essentially
depends on $M$.
In the $M\to\infty$ limit we have
\be
\tilde{\cal C}(u)=H^\perp(u^{-2})=
\sum_{n=0}^\infty u^{-2n} h^{\perp}_n,
\label{C}
\ee
where $h^{\perp}_n$ is the adjoint to the operator of multiplication
by $h_n$ in the whole space $\La$
(cf.\ \cite[Ex. I.5.3, Ex. I.5.29]{Mac}).

\subsection{Vertex operators and enumeration of plane partitions}
\label{sec:vertex}

\begin{lemma}
In the $M\to\infty$ limit, 
the operator $\tilde{\cal B}(u)$ 
has the following vertex operator representation:
\be
\tilde{\cal B}(u)=\exp\left(\sum_{k=1}^\infty\frac{u^{2k}}{k}\alpha_{-k}\right),
\label{vertex}
\ee
where $\alpha_{-k}$, $k=1,2,\ldots$, are the free boson operators.
\end{lemma}

\begin{proof}
By the well-known formula \cite[I.2.10]{Mac},
$$
\frac{d}{dt}\log H(t)=P(t),
$$
where $P(t)=\sum_{k=1}^\infty t^{k-1}p_k$ is the generating
function of the
Newton power sums $p_k$. Thus we have
$$
H(t)=\exp\left(\int P(t)\right)
=\exp\left(\sum_{k=1}^\infty\frac{t^k}{k}p_k\right).
$$
On the other hand, it is well known 
that in the realization of the Fock 
space as the algebra of symmetric functions, the free boson operator
$\alpha_{-k}$ corresponds to the multiplication by $p_k$,
so that (\ref{vertex}) follows by Proposition~\ref{prop:B}.
\end{proof}

Note that the vertex operator in the right-hand side of (\ref{vertex})
is exactly the operator used by 
Okounkov and Reshetikhin  \cite{OkResh} in connection with 
computing the correlation functions
of plane partitions. Namely, in the notation of \cite{OkResh},
$$
\tilde B_0(q^\frac j2)=\Gamma_+(\phi_j),\qquad
\mbox{where}\qquad
\phi_j(z)=\phi_{\rm 3D}[j](z)=\frac{1}{1-q^jz}.
$$
In particular, in the symmetric function realization,
the vertex operator associated with the Schur process describing
plane partitions is just the operator of multiplication by the generating
function of complete symmetric functions:
$$
\Gamma_+(\phi_{\rm 3D}[j])=H(q^j).
$$

In view of (\ref{C}), 
the $M\to\infty$ limit of the regularized annihilation operators
$\tilde{\cal C}(v)=v^{-M}{\cal C}(v)$ has the following vertex representation:
$$
\tilde {\cal C}(v)=
\exp\left(\sum_{k=1}^\infty\frac{v^{-2k}}{k}\alpha_{k}\right).
$$

\subsection{Commutation relations for the ``truncated'' operators}
\label{sec:commrel}

Using the commutation relations for vertex operators
(see, e.g., \cite[(14.10.12)]{Kac} or \cite[(11)]{OkResh}), 
one can easily obtain the well-known
commutation relation \cite[Ex. I.5.29, (2)]{Mac} for 
the operators $H$ and $H^\perp$
in the whole
algebra of symmetric functions $\La$:
\be
H^\perp(u)H(v)=\frac{1}{1-uv}H(v)H^\perp(u).
\label{comminf}
\ee

However, in
the subspace $\La_M$  generated by Schur functions whose
diagrams have at most
$M$ columns, the vertex representation (\ref{vertex})
and the commutation relation (\ref{comminf}) are no longer
valid. Nevertheless, we can obtain the commutation relation for 
$H_M(v)=\sum_{k=0}^Mv^kh_k$ and $H_M^\perp(v)$
in $\La_M$ using the QISM 
machinery \cite{KBI} and the above 
interpretation of the phase model 
in terms of symmetric functions.
Namely, the bilinear equation (\ref{Tbil})
implies, in particular, that
$$
D(u)B(v)=\frac{u^2}{u^2-v^2}B(v)D(u)-\frac{uv}{u^2-v^2}B(u)D(v).
$$
Using Proposition~\ref{prop:B} and Lemma~\ref{l:ABCD},
we obtain
\be
H^\perp_M(u)H_M(v)=\frac{1}{1-uv}\left[H_M(v)H^\perp_M(u)-
(uv)^{M+1} H_M(u^{-1})H_M^\perp(v^{-1})\right].
\label{commfin}
\ee
We see that in the formal $M\to\infty$ limit 
with $|uv|<1$, relation (\ref{commfin})
reduces to (\ref{comminf}).

Expanding both sides of (\ref{commfin}) into power series in $u,v$ and equating
coefficients, we obtain the following commutation relations
in $\La_M$:
$$
h^{\perp,M}_m h_n=\sum_{i=0}^{\min\{m,n\}}h_{n-i}h^{\perp,M}_{m-i}-
\sum_{i=0}^{\min\{m,n\}-1}h_{M+1-m+i}h^{\perp,M}_{M+1-n+i}.
$$

\smallskip\noindent{\bf Examples.}
For $M=1$, we obtain  the relation 
$h_1^{\perp,1} h_1=1$ in $\La_1$. Indeed, $\La_1$ is generated by Schur functions
with one-column diagrams,
so that the operators $h_1$ and $h_1^{\perp,1}$ correspond to
adding and removing one cell, respectively,
i.e., they are one-sided shifts.

For $M=2$, we obtain
$$
h^{\perp, 2}_1h_1=h_1h^{\perp, 2}_1+1-h_2h^{\perp, 2}_2,\qquad
h^{\perp, 2}_1h_2=h_1,\qquad h^{\perp, 2}_2h_1=h_1^{\perp, 2},\qquad
h^{\perp, 2}_2h_2=1.
$$

\section{$q$-Boson model and Hall--Littlewood functions}
\label{sec:q-boson}

\subsection{$q$-Boson model}
\label{sec:q}

The phase model considered in the previous section is a particular case of
the so-called {\it $q$-boson} model \cite{BBT, BIK2}. 

Let $q$ be a nonnegative parameter.
Consider the {\it $q$-boson algebra}
generated by three operators $B,\Bip, N$ with 
commutation relations
$$
[N,B]=-B,\qquad [N,\Bip]=\Bip,\qquad [B,\Bip]=q^{2N}.
$$

Denote
$$
[n]=\frac{1-q^{2n}}{1-q^2},\qquad [n]!=\prod_{j=1}^n[j].
$$
The standard realization of the $q$-boson algebra in the  Fock 
space $\cal F$ looks as follows:
$$
\Bip|n\rangle=[n+1]^{\frac12}|n+1\rangle, \qquad
B|n\rangle=[n]^{\frac12}|n-1\rangle,\; B|0\rangle=0,\qquad
N|n\rangle=n|n\rangle.
$$
However, it will be more convenient to use another realization,
namely,
\be
\Bip|n\rangle=[n+1]|n+1\rangle, \qquad
B|n\rangle=|n-1\rangle,\; B|0\rangle=0,\qquad
N|n\rangle=n|n\rangle.
\label{realizq1}
\ee
For the operators $B$ and $\Bip$ to be adjoint to each other, we should
normalize the Fock vectors so that
\be
<n|n>^2=\frac{1}{[n]!}.
\label{normalization}
\ee
Yet another realization of the $q$-boson model in the Fock space is
given by the formula
\be
\Bip|n\rangle=|n+1\rangle, \qquad
B|n\rangle=[n]|n-1\rangle,\; B|0\rangle=0,\qquad
N|n\rangle=n|n\rangle
\label{realizq2}
\ee
with the normalization
\be
<n|n>^2=[n]!.
\label{normaliz2}
\ee

One can easily see that the phase model is the particular case of the $q$-boson
model corresponding to $q=0$. If $q\to1$, then the operators $B$ and $\Bip$
turn into the canonical 
free boson operators $b$ and $b^\dagger$, respectively,
satisfying the commutation relation
$[b,b^\dagger]=1$.

Now we apply the same scheme as we have used for the phase model in 
Sec.~\ref{sec:phase}:
fix the number of  sites $M$, consider the tensor product 
${\cal F}={\cal F}_0\otimes{\cal F}_1\otimes\ldots\otimes{\cal F}_{M}$
of $M+1$ copies  ${\cal F}_i$, $i=0,\ldots,M$, of the
one-dimensional Fock space,
and denote by $B_i,\Bip_i, N_i$ the operators that act as
$B,\Bip, N$, respectively, in the $i$th space, and identically
in the other spaces. It will be convenient to use the 
realization~(\ref{realizq1}) of the $q$-boson algebra for $i=1,\ldots,M$
and the realization~(\ref{realizq2}) for $i=0$.

Note that in view of (\ref{normalization}) and (\ref{normaliz2})
the squared norms of the basis $N$-particle vectors (\ref{Npart}) are equal to
\be
\|\psi_{n_0,\ldots,n_M}\|^2=\frac{[n_0]}{\prod_{j=1}^M[n_j]!}.
\label{psinorms}
\ee

The Hamiltonian of the $q$-boson model has the form
$$
H=-\frac12\sum_{n=0}^M(\Bip_nB_{n+1}+B_n\Bip_{n+1}-2N_n),
$$
with the periodic boundary conditions: $M+1\equiv 1$.

The {\it $L$-matrix} for the $q$-boson model is given by
\be
L_0(u)=\left(\begin{array}{cc}
u^{-1}I &\Bip_0\\
(1-q^2)B_0 & uI
\end{array}\right),\qquad
L_n(u)=\left(\begin{array}{cc}
u^{-1}I & (1-q^2)\Bip_n\\
B_n & uI
\end{array}\right),\qquad n=1,\ldots,M.
\label{L-q}
\ee
This $L$-matrix satisfies the bilinear equation (\ref{Lbil})
with the $R$-matrix
\be
R(u,v)=\left(\begin{array}{cccc}
f(v,u) & 0 & 0 & 0\\
0 & g(v,u) & q^{-1} & 0\\
0 & q & g(v,u) & 0\\
0 & 0 & 0 & f(v,u)
\end{array}\right),
\label{R-qboson}
\ee
where
\be
f(v,u)=\frac{q^{-1}u^2-qv^2}{u^2-v^2},\qquad g(v,u)=\frac{uv}{u^2-v^2}(q^{-1}-q).
\label{fg-q}
\ee
Note that the 
R-matrix (\ref{R-phase}), (\ref{fg-phase}) of the phase
model is obtained from the 
R-matrix of the $q$-boson model as 
the renormalized $q\to0$ limit:
$R_{\rm phase}=\lim_{q\to0}qR_{q\mbox{\rm\scriptsize -boson}}$.

Denote by 
$$
T(u)=L_M(u)\ldots L_0(u)=\left(\begin{array}{cc}
A(u) & B(u)\\
C(u) & D(u)
\end{array}\right)
$$
the monodromy matrix of the $q$-boson model. It satisfies the bilinear equation 
(\ref{Tbil}) with the $R$-matrix (\ref{R-qboson}), (\ref{fg-q}).

\subsection{Hall--Littlewood functions}
\label{sec:HL}

In this section, we give a brief account of basic facts
related to the Hall--Littlewood symmetric functions;
for details, see \cite[Ch.~III]{Mac}.

The Hall--Littlewood symmetric functions with parameter $t\ge0$,
indexed by Young diagrams $\la$,
are defined, for example, as follows. First, for a finite
number $n\ge l(\la)$ of variables $x_1,\ldots,x_n$, set
$$
P_\la(x_1,\ldots,x_n;t)=\frac{1}{v_\la(t)}\sum_{w\in{\mathfrak S}_n}
w\left(x_1^{\la_1}\ldots x_n^{\la_n}\prod_{i<j}\frac{x_i-tx_j}{x_i-x_j}\right),
$$
where ${\mathfrak S}_n$ is the symmetric group of degree $n$ 
acting by permutations of variables and
$$
v_\la(t)=\prod_{i\ge0}v_{n_i(\la)}(t),\qquad 
v_n(t)=\prod_{i=1}^n\frac{1-t^i}{1-t}.
$$
Then observe that for any diagram $\la$ with $l(\la)\le n$,
we have $P_\la(x_1,\ldots,x_n;t)=P_\la(x_1,\ldots,x_n,0;t)$,
so that we can define a symmetric function $P_\la(x;t)$ of
infinitely many variables with coefficients in 
${\mathbb Z}[t]$ 
as the inductive limit of $P_\la(x_1,\ldots,x_n;t)$
with respect to the projections sending the last variable to $0$.
The functions $P_\la(x;t)$ form a ${\mathbb Z}[t]$-basis of 
the algebra $\La[t]$ of symmetric functions with coefficients in ${\mathbb Z}[t]$.
They interpolate between the Schur functions $s_\la$ and the
monomial symmetric functions $m_\la$:
\be
\label{HLschurs}
P_\la(x;0)=s_\la(x),\qquad
P_\la(x;1)=m_\la(x).
\ee

It is convenient to introduce another family of symmetric functions
$Q_\la(x;t)$ that are scalar multiples of $P_\la(x;t)$.
Namely, we set
$$
Q_\la(x;t)=b_\la(t)P_\la(x;t),
$$
where 
$$
b_\la(t)=\prod_{i\ge1}\phi_{n_i(\la)}(t),\qquad
\phi_n(t)=(1-t)(1-t^2)\ldots(1-t^n).
$$
In the case $q=0$ we have $Q_\la(x;0)=P_\la(x;0)=s_\la(x)$.

Now set
$$
q_r(x;t)=Q_{(r)}(x;t)=(1-t)P_{(r)}(x;t),\quad r\ge1;\qquad
q_0(x,t)=1.
$$
The generating function for $q_r$ is equal to
\be
Q(u)=\sum_{r=0}^\infty q_r(x;t)u^r=\prod_i\frac{1-x_itu}{1-x_iu}=
\frac{H(u)}{H(tu)},
\label{Q}
\ee
where $H(u)$ is the generating function of the complete symmetric functions.
In particular, 
\begin{eqnarray}
q_r(x,0)&=&h_r(x),\label{888}\\
q_r(x;1)&=&0 \quad\mbox{ for }  r\ge1.
\label{999}
\end{eqnarray} 
Let
\be
q_\la(x;t)=\prod_{i\ge0}q_{\la_i}(x;t).
\label{qla}
\ee
The symmetric functions $q_\la(x;t)$ form a ${\mathbb Q}[t]$-basis
of $\La[t]$. 

For $t\ne1$,  introduce a scalar product in 
$\La[t]$ by requiring that the
bases $\{q_\la\}$ and $\{m_\la\}$ be dual to each other:
$$
\langle q_\la(x;t), m_\la(x)\rangle=\delta_{\la\mu}.
$$
Then the bases $\{P_\la\}$ and $\{Q_\la\}$ are also dual:
$$
\langle P_\la(x;t), Q_\la(x;t)\rangle=\delta_{\la\mu},
$$
so that the squared norm of the Hall--Littlewood function $P_\la(x;t)$ equals
\be
\langle P_\la(x;t), P_\la(x;t)\rangle=\frac{1}{b_\la(t)}.
\label{HLnorms}
\ee

Observe that in the case $t=0$ this scalar product reduces to the
standard scalar product in $\La$ (with respect to which the Schur
functions form an orthonormalized basis).

The generalization of the Cauchy identity to the case of Hall--Littlewood functions
looks as follows:
\be
\prod_{i,j}\frac{1-tx_iy_j}{1-x_iy_j}=\sum_\la P_\la(x;t)Q_\la(y;t)=
\sum_\la b_\la(t)P_\la(x;t)P_\la(y;t).
\label{HL-Cauchy}
\ee
There is also a generalization of the Pieri formula. Namely, 
\be
P_\mu q_r=\sum_{\la:\la\setminus\mu\in{\cal H}_k}\phi_{\la\setminus\mu}(t)P_\la,
\label{HL-Pieri}
\ee
with
\be
\phi_{\la\setminus\mu}(t)=\prod_{i\in I}(1-t^{n_i(\la)}),
\label{phi}
\ee
where $\theta=\la\setminus\mu$ and 
$I=\{i:\theta'_i=1,\;\theta'_{i+1}=0\}$ (recall that $\theta'_i$ is the length
of the $i$th column of the skew diagram $\theta$, which in the case of
a horizontal strip can be equal to $0$ or $1$).

\subsection{Realization of the $q$-boson model in the algebra of symmetric functions}
\label{sec:qHL}

We will follow the same scheme as was used in Sec.~\ref{sec:symmphase}
for the phase model.

With a basis vector (\ref{Npart})
we associate
the Hall--Littlewood function $P_\la(x;q^2)$ with the diagram
determined by the occupation numbers:
$$
\bigotimes_{j=0}^M|n_j\rangle_j\leftrightarrow P_\la(x;q^2),\qquad \la=1^{n_1}2^{n_2}\ldots.
$$
Note that in view of (\ref{psinorms}) and (\ref{HLnorms})
this correspondence is not an isometry.

Set ${\cal B}(u)=PB(u)P$,
where $P$ is the projection to the positive-energy subspace.
Denote by $\La_M[q^2]$ the subspace in $\La[q^2]$ spanned by
Hall--Littlewood functions $P_\la(x;q^2)$ with diagrams having
at most $M$ columns.

\begin{proposition} 
Let ${\cal B}(u)=u^{-M}\tilde{\cal B}(u)$. 
The operator $\tilde{\cal B}(u)$ acts in $\La_M[q^2]$
as the operator of multiplication by $Q_M(u^2)$,
where $Q_M(t)=\sum_{k=0}^M t^kq_k(x;q^2)$.
\label{prop:qB}
\end{proposition}

\begin{proof}
Arguing as in the proof of Proposition~\ref{prop:B}, we see that 
$$
{\cal B}_k(u)P_\mu(x;q^2)=\sum_{\la:\,\la\setminus\mu\in{\cal H}_k}
c(\mu,\la)P_\la(x;q^2),
$$
but the coefficient $c(\mu,\la)$ is no longer equal to 1.
However, we can easily compute it. Indeed, the coefficient arises
from applying the creation operators $\Bip_j$ with $j\ge1$. 
Namely, if $\B_{\eps_M,\ldots,\eps_0}s_\mu=s_\la$, then $c(\mu,\la)$
is the product of the factors 
$(1-q^2)[n_i(\mu)+1]=1-q^{2(n_i(\mu)+1)}$ 
over all $i\ge1$ such that $\eps_i=1$. But the latter condition is equivalent to
$n_i(\la)=n_i(\mu)+1$, or $\theta'_i=1$,
$\theta'_{i+1}=0$, i.e., the product is over all $i$ belonging to the set $I$ 
in the notation of (\ref{phi}).
Thus we have
$$
c(\mu,\la)=\prod_{i\in I}(1-q^{2n_i(\la)})=\phi_{\la\setminus\mu}(q^2),
$$
and the proposition follows by the Pieri-type formula~(\ref{HL-Pieri})
for the Hall--Littlewood functions.
\end{proof}

\smallskip\noindent{\bf Remark.} As in the case of complete symmetric functions, 
we may regard the truncated generating function $Q_M(t)$ as the full
generating function $Q(t)$ under an appropriate specialization:
\be
Q_M(t)=Q(t)\left|_{q_{M+1}=q_{M+2}=\ldots=0}.\right.
\label{specHL}
\ee

\begin{corollary} There is a well-defined $M\to\infty$ limit
of the operator $\tilde\B(u)$. In the realization of the $q$-boson
model in the algebra of symmetric functions,
it is the operator of multiplication by $Q(u^2)=\frac{H(u^2)}{H(q^2u^2)}$.
\end{corollary}

\begin{proposition}
Let $|\Psi_N(u_1,\ldots,u_N)\rangle=\prod_{j=1}^NB(u_j)|0\rangle$.
Then
$$
|\Psi_N(u_1,\ldots,u_N)\rangle=
\sum_{\la}Q_\la(u_1^2,\ldots, u_N^2;q^2)\bigotimes_{j=0}^M|n_j\rangle_j,
$$
where the sum is over all Young diagrams $\la$ with at most $N$ rows
and at most $M$ columns.
\end{proposition}
\begin{proof} The proof is similar to that of Proposition~\ref{prop:schurs}
and uses Proposition~\ref{prop:qB}, the formula~(\ref{Q}) for the
generating function of $q_k(x;t)$, and the Cauchy-type identity~(\ref{HL-Cauchy})
for the Hall--Littlewood functions. The restriction on diagrams $\la$
follows  from the following facts: (a) $P_\la(x;t)=0$ if the number of nonzero
variables $x_i$ is less than $l(\la)$ (this easily follows from the definition
of the Hall--Littlewood functions) and
(b) the transition matrix from the basis
$\{Q_\la\}$ to the basis $\{q_\la\}$ is strictly lower triangular
(\cite[III.2.16]{Mac}), so that the specialization~(\ref{specHL})
implies, in view of~(\ref{qla}), that $Q_\la=0$ unless $\la_1\le M$.
\end{proof}

\smallskip\noindent{\bf Remark.}
In the case $q=0$, the results of this section reproduce those
of Sec.~\ref{sec:symmphase} in view of~(\ref{HLschurs}), (\ref{888}). 

In the case $q=1$, the $L$-matrix (\ref{L-q}) degenerates into a 
lower triangular matrix
for $n=1,\ldots,M$ and to an upper triangular matrix for $n=0$,
so that $B(u)=u^{-M}\Bip_0$, whence $\tilde{\cal B}(u)=1$,
in accordance with~(\ref{999}).

\bigskip
The author is grateful to N.~M.~Bogoliubov for introducing into the 
quantum inverse scattering method and $q$-boson model, and to
A.~M.~Vershik for many useful discussions.

\end{document}